\documentclass[aps,prb,twocolumn,superscriptaddress,showpacs,preprintnumbers]{revtex4-1}
\usepackage{graphicx}

\begin{document}

\preprint{Published in Phys. Rev. B 89, 115105 (2014) \copyright American Physical Society}

\title{Relaxation Dynamics of Photoexcited Charge Carriers at the Bi(111) Surface}


\author{Christopher Bronner}
    \email{bronner@uni-heidelberg.de}

\author{Petra Tegeder}
\affiliation{Ruprecht-Karls-Universit\"at Heidelberg, Physikalisch-Chemisches Institut, Im Neuenheimer Feld 253, 69120 Heidelberg, Germany}
\affiliation{Freie Universit\"at Berlin, Fachbereich Physik, Arnimallee 14, 14195 Berlin, Germany}


\date{published in Phys. Rev. B on 5 March 2014}

\begin{abstract}
Bi possesses intriguing properties due to its large spin-orbit coupling, e.g. as a constituent of topological insulators. While its electronic structure and the dynamics of electron-phonon coupling have been studied in the past, photo-induced charge carriers have not been observed in the early phases of their respective relaxation pathways. Using two-photon photoemission (2PPE) we follow the de-excitation pathway of electrons along the unoccupied band structure and into a bulk hole pocket. Two decay channels are found, one of which involves an Auger process. In the hole pocket, the electrons undergo an energetic stabilization and recombine with the corresponding holes with an inverse rate of 2.5~ps. Our results contribute to the understanding of the charge carrier relaxation processes immediately upon photo-excitation, particularly along the $\Gamma T$-line where the electron dynamics have not been probed with time-resolved 2PPE so far.
\end{abstract}


\maketitle


\section{Introduction}

The electronic structure of Bi and especially its (111)-surface have received much attention in the past years due to their intriguing properties.\cite{Hofmann2006} Being a relatively heavy element, Bi exhibits a large spin-orbit coupling which causes a large spin-polarization of the Bi(111) surface states\cite{{Koroteev2004},{Ohtsubo2012a}} and makes this material interesting for spintronics.\cite{{Datta1990},{Koga2002}} Additionally, Bi is a constituent of many topological insulators such as $\text{Bi}_{1-x}\text{Sb}_{x}$ (which is merely a partially substituted Bi crystal)\cite{Hsieh2008} and layered materials, e.g. $\text{Bi}_{2}\text{Se}_{3}$ and $\text{Bi}_{2}\text{Te}_{3}$.\cite{Zhang2009}

\begin{figure}[htbp]
\includegraphics{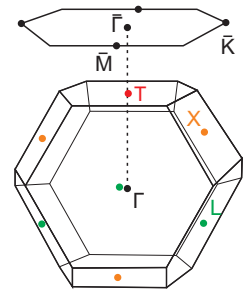}
\caption{(Color online) Brillouin zone of the rhombohedral Bi lattice together with the surface Brillouin zone and high-symmetry points.\label{fig:Fig0-BrillouinZone}}
\end{figure}

Using electron spectroscopies, the band structure and the dispersive surface states of Bi(111) have been studied extensively in the past. The three occupied \emph{p}-bands all disperse with negative dispersion parallel to the surface\cite{{Thomas1999},{Tanaka1999}} and the highest lying band crosses the Fermi level at the $T$-point of the bulk Brillouin zone (see Fig. \ref{fig:Fig0-BrillouinZone}), forming a hole pocket.\cite{Ast2004} Bulk electron pockets have not been observed in photoemission experiments so far but have been predicted to occur at the $L$-points (analogously to the hole pockets at the $T$-points).\cite{{Golin1968},{Liu1995}} These hole and electron pockets, respectively, are not very extended in reciprocal space but they constitute the entire bulk Fermi surface of Bi which is why Bi is a semi-metal. In the regime above the Fermi level we have recently reported observation of all three unoccupied $p$-bands using two-photon photoemission (2PPE) and their dispersion along the $\bar\Gamma\bar K$-line of the surface Brillouin zone (SBZ).\cite{Bronner2013} Besides these three bands, we also observed a signature of the bulk hole pocket at an energy of $0.11~\text{eV}$ with respect to the Fermi level. Throughout the paper we will adapt the band index notation of Ast and H\"{o}chst\cite{Ast2004}, i.e. the occupied $p$-bands are labeled as band 3-5 (with decreasing energy difference to the Fermi level) and the unoccupied $p$-bands are named band 6-8 (see Fig. \ref{fig:Fig2-ElectronRelaxation}). While all these features originate in the bulk band structure, Bi(111) exhibits a number of surface states and resonances, two of which are located within the occupied $p$-bands and generally exhibit a negative dispersion\cite{{Ast2002},{Ast2003},{Jezequel1986},{Patthey1994}}. Due to their sixfold symmetry they can more accurately be described as surface resonances rather than surface states. Another surface state lies very close to the Fermi level and manifests itself in two surface bands which form a pronounced electron pocket\cite{{Ast2001},{Koroteev2004},{Ast2003},{Ohtsubo2012a}} at the $\bar\Gamma$-point and the $\bar M$-point\cite{{Hengsberger2000},{Ast2003}} as well as sixfold droplet-shaped hole pockets\cite{{Hengsberger2000},{Ast2001},{Koroteev2004},{Ast2003},{Ohtsubo2012a}} along the $\bar\Gamma\bar M$-lines of the SBZ. By virtue of these pockets, this spin-split surface state constitutes most of the Fermi surface of Bi(111) which makes the surface more metallic than the semi-metallic bulk. Furthermore, close to the vacuum level, unoccupied image potential states (IPS) have been observed using 2PPE.\cite{{Muntwiler2008},{Bronner2013}} With this comprehensive background of studies on the occupied and unoccupied electronic structure of the Bi(111) surface, research currently focusses on the dynamics of the electronic system following optical excitation by ultrashort laser pulses.

Several time-resolved studies have been carried out on photo-excited Bi using THz spectroscopy\cite{Timrov2012}, time-resolved 2PPE\cite{{Papalazarou2012},{Faure2013}} or pump-probe reflectivity measurements.\cite{Zeiger1992} Particularly the relaxation pathway of the out-of-equilibrium electronic system as well as quantification of the decay dynamics are of interest. Generally, the electronic system is thought to first thermalize by fast electron-electron scattering, whereupon (on a longer timescale) the electron-hole pairs recombine via coupling to phonons. Timrov \emph{et al.} used laser pulses with a photon energy of 1.6~eV to excite the electronic system and probed the relaxation dynamics with THz spectroscopy. A peculiar, non-monotonic behavior of the plasma frequency has been explained by a two-step relaxation process: in the first step, with a time constant of 0.6~ps, holes decay from a local maximum of band 4 near the $\Gamma$-point toward the hole pocket at the $T$-point where they recombine with excited electrons in a second step which has a time constant of 4.0~ps.\cite{Timrov2012} In a time-resolved 2PPE experiment, it has been found that the energy of a bulk band of Bi oscillates with a frequency of 2.97~THz, in coincidence with the frequency of the $A_{1g}$ phonon mode which is essentially a lattice distortion along the [111] axis. This electron-phonon coupling is associated with the slow decay component on the picosecond timescale.\cite{Papalazarou2012} Note, that close to the Fermi level, the phase-space for electronic relaxation has a much smaller volume due to the low density of states of this semi-metal which is the reason for relatively slow relaxation processes on the picosecond timescale. More recently it was found that excited electrons accumulate in the unoccupied branch of the surface state band within a few hundred femtoseconds upon photoexcitation, i.e. after electronic thermalization. The resulting electron-hole plasma situated along the $\bar\Gamma\bar M$-line of the SBZ then decays on the timescale of 5-6~ps via electron-phonon coupling to the $A_{1g}$ mode.\cite{{Ohtsubo2012a},{Faure2013}} While these studies provide an extensive insight into the dynamics following photoexcitation, no direct evidence of the initially excited electrons is observed in the first phase after the pump pulse.

In the present contribution we use time-resolved 2PPE (TR-2PPE) at the $\bar\Gamma$-point of Bi(111). Besides short-lived IPS we observe photo-excited electrons in band 6 which decay from the $\Gamma$-point into the bulk hole pocket at the $T$-point via two different channels. Simultaneously, the photo-hole relaxes toward the hole pocket and we observe recombination of the electron-hole pairs with a time constant of 2.5~ps.

\section{Experimental methods}

Two-photon photoemission (2PPE) is a pump-probe technique in which the pump laser pulse excites an electron from an occupied electronic state to an unoccupied state. The probe pulse then energetically lifts the electron above the vacuum level $E_{\text{vac}}$ and the emitted photoelectron can be detected in a time-of-flight electron spectrometer. Pump and probe pulse can be delayed with respect to each other in order to obtain information about the evolution of an electronic state on an ultrashort timescale. Features in a 2PPE spectrum may arise not only due to photoemission from intermediate states as described above but also from direct two-photon excitation of occupied states via a virtual intermediate state. Due to this ambiguity and because one has to distinguish between electrons probed with one pulse or the other (depending on the sign of the time delay), we display the 2PPE spectra as a function of the final state energy $E_{\text{Final}}-E_{\text{F}}=E_{\text{kin}}+\Phi$ which is referenced to the Fermi level $E_{\text{F}}$ and which corresponds to the energy of the photoelectron $E_{\text{kin}}$ plus the work function $\Phi$ of the sample.

Femtosecond laser pulses with a wavelength of 800~nm were created in a Ti:Sapphire oscillator and amplified to reach pulse energies on the order of $\mu\text{J}$. The resulting ultrashort pulses were converted using an optical parametric amplifier (OPA) which yielded pulses in the visible spectrum which in turn could be frequency-doubled into the UV range using a BBO crystal. The OPA furthermore provides an output of the second harmonic of the original pulses which has an energy of 3.08~eV. In the experiments reported in the present paper we used both the visible and the 3.08~eV pulses in combination with the UV pulses for two-color 2PPE (2C-2PPE). All beams were p-polarized.

The sample is mounted in ultrahigh vacuum and on a cryostat equipped with resistive heating, thus allowing temperature control from 90~K to the annealing temperature of Bi. The chamber is equipped with an ion source for sputtering, facilities to perform low energy electron diffraction (LEED) for sample characterization and a time-of-flight spectrometer. The Bi(111) single crystal was prepared by sputtering (900~V) and annealing (410~K, 10~min) and the quality of the surface was checked with LEED and by comparison of the IPS and the work function with results from the literature.\cite{Muntwiler2008} The setup is described in great detail elsewhere.\cite{{Hagen2010},{Bronner2012},{Tegeder2012}}

\section{Results}

\begin{figure*}[htbp]
\includegraphics{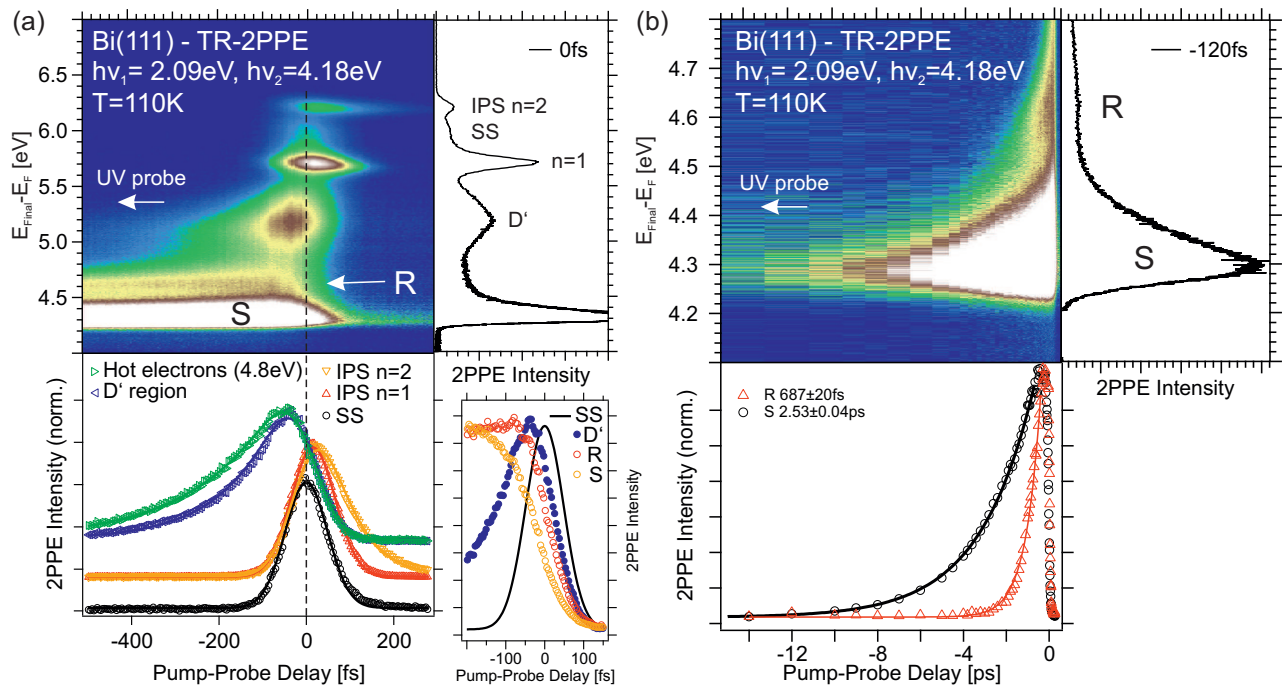}
\caption{(Color online) (a) TR-2PPE data shown in a false-color representation versus final state energy and time delay, where negative delays correspond to the situation in which the UV pulse is the probe pulse. On the right side of the figure, a spectral cut at the time-overlap of the laser pulses (zero delay) is shown and below, cross-correlation (XC) cuts are shown with fits (see text) for the surface state (SS), the two IPS and for feature D' as well as a spectral region at slightly lower energies than D', namely at 4.8~eV, corresponding to hot electron contributions. The XC traces are normalized to the maximum intensity. In the lower right corner, XC traces are also shown for features R and S. (b) The same TR-2PPE measurement on a longer timescale. The false-color plot shows a magnified region above the Fermi level, the spectral cut on the right shows peaks R and S at a time delay of 120~fs and in the graph below the false-color plot, the normalized XC curves for R and S are shown together with exponential decay fits.\label{fig:Fig1-TR2PPE}}
\end{figure*}

One of the most important advantages of 2PPE is the ability to study the dynamics of charge carriers on a femtosecond timescale which is done by varying the pump-probe delay. Fig. \ref{fig:Fig1-TR2PPE}a shows the result of such a TR-2PPE experiment in a false-color representation. Vertical cuts to such a plot yield 2PPE spectra at a fixed pump-probe delay while horizontal cuts give the so-called cross-correlation (XC) curves. The temporal resolution is determined by the pulse width of the laser pulses, for which the auto-correlation (AC) is a measure. The AC is the temporal convolution of the pump and the probe pulse and easily accessible in experiment as the XC curve of an occupied state. The full width at half maximum of a single laser pulse is here determined to be 81~fs from the surface state peak at $E_{\rm{Final}}-E_{\rm{F}}=6.0~{\rm{eV}}$. This peak also provides the origin for the femtosecond time axis (zero delay).\\
At higher energies in the 2PPE spectrum, two distinct features at final state energies of 5.7~eV and 6.2~eV are observed which show an asymmetry in intensity to positive delays which indicates that they are probed with the visible photons in this experiment. This allows an immediate identification of those two peaks to be caused by photoemission from the $n=1$ and $n=2$ IPS.\cite{{Muntwiler2008},{Bronner2013}} One characteristic of these states is their lifetime $\tau$. If we assume a simple exponential decay of the electrons in these IPS, i.e. $N(t)\propto\exp(-t/\tau)$, the lifetime can be obtained from a fit of the convolution of the state's time-dependent population $N(t)$ with the temporal profile of both laser pulses.\cite{Weinelt2002} Such fits are shown together with the XC curves in Fig. \ref{fig:Fig1-TR2PPE}a. The use of a convolution becomes increasingly important as the lifetimes of the observed states decrease since, in the range of several 10~fs, the lifetimes are much shorter than the pulse duration. For the more intense $n=1$ IPS we find a lifetime of $\tau_{1}=12~{\rm{fs}}$ while the higher-lying $n=2$ state possesses a longer lifetime of $\tau_{2}=65~{\rm{fs}}$ in agreement with previous 2PPE measurements.\cite{Muntwiler2008}
At lower energies, another feature D' is observed at 5.2~eV in the 2PPE spectra. It has previously been assigned to the unoccupied \emph{p}-band 6 near the $\Gamma$-point.\cite{Bronner2013} The XC curve of D' shows a strong asymmetry toward negative delays (i.e. it is probed with the UV photons) which can be fitted in the same manner as described above for the IPS. The lifetime obtained in this way is $\tau_{D'}=125~{\rm{fs}}$, however peak D' may be located on a hot electron (HE) background in the spectrum. Fitting the XC curve of the region around 4.8~eV yields a lifetime of $\tau_{\text{HE}}=187~{\rm{fs}}$. Considering the energy-dependence of hot electron lifetimes above the Fermi level which can be described by the Fermi liquid theory\cite{Ogawa1997} by $\tau_{\text{HE}}\propto(E-E_{\rm{F}})^{-2}$, we can estimate the lifetime of hot electrons in the energetic region where D' is located to be around 70~fs. While this is a very rough estimate it demonstrates that the asymmetry of D' may to a large fraction be owed to hot electrons and we simply conclude that the lifetime of D' is less than 125~fs.

At binding energies $E_{\rm{R}}=0.44\pm0.02~{\rm{eV}}$ and $E_{\rm{S}}=0.11\pm0.03~{\rm{eV}}$, respectively, two additional features labeled R and S are observed in the spectrum. Feature R is due to photoemission from band 6 near the $T$-point whereas peak S corresponds to the bulk hole pocket which is formed by band 5.\cite{Bronner2013} Note, that both features D' and R result from the same band which is probed in two different points in the Brillouin zone.
States R and S both exhibit a lifetime on the picosecond timescale, as can already be seen in Fig. \ref{fig:Fig1-TR2PPE} by the naked eye. Let us, however, first consider the population dynamics which occur on a much shorter timescale just after the visible pump pulse (see lower right inset in Fig. \ref{fig:Fig1-TR2PPE}a). There, the XC curves of D', R and S are shown together for small negative time delays. The population of D' occurs fastest at zero delay, i.e. when the pump pulse is most intense. This behavior is expected for a direct population process in which electrons are optically excited from an occupied state to the observed unoccupied state. There are however other (indirect) population pathways, in which higher lying electronic states are excited by the laser pulse and the excited electron relaxes to the observed state on an ultrafast timescale. This should in general be evident by a delayed population buildup, unless the ultrafast relaxation into the observed state is faster than the temporal resolution. Considering the band structure (see below) we believe that the latter is the case for feature D'. The rising edge of the XC curves of R and S on the other hand is not observed at zero delay but at a delay of 20~fs and 50~fs (values for half buildup), respectively, i.e. after the visible pulse which indicates indirect population of these states.
In Fig. \ref{fig:Fig1-TR2PPE}b, the same TR-2PPE measurement is shown but for higher pump-probe delays up to 14~ps. Since here, the pulse duration is negligible compared to the lifetimes of R and S, we used only a simple exponential decay function to obtain the lifetimes, namely $\tau_{R}=687\pm20~{\rm{fs}}$ and $\tau_{S}=2.53\pm0.04~{\rm{ps}}$. These values are quite large as electrons in bulk bands usually decay on the order of a few femtoseconds due to the high available phase space. Note that we observe two features in the spectrum associated with band 6, one close to the $\Gamma$-point (D'), one close to the $T$-point (R). However, the lifetimes of both excited states have drastic differences. While the feature from the $\Gamma$-point (D') has a lifetime of less than 125~fs (probably even less since the XC curve of D' is influenced by hot electrons), the lifetime of the electrons in this band 6 is strongly increased to 687~fs near the $T$-point where a true local minimum of the band structure exists.\cite{Timrov2012}

\subsection{Electron relaxation}

\begin{figure}[htbp]
\includegraphics{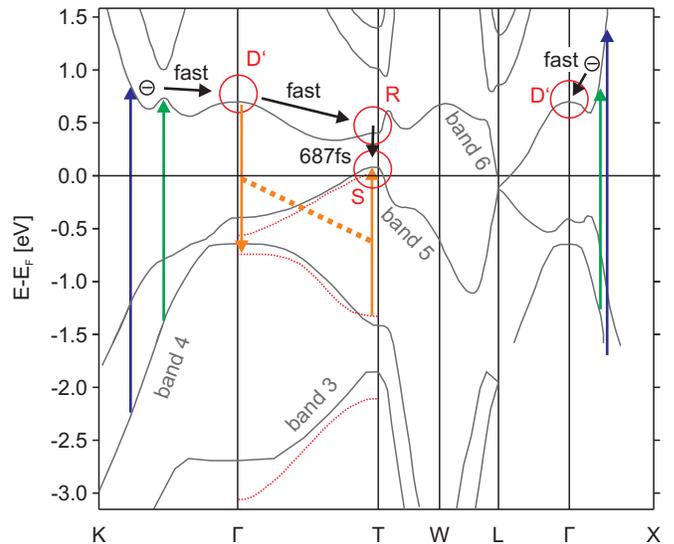}
\caption{(Color online) Band structure of Bi as calculated in ref. \cite{Timrov2012} (grey lines) and proposed decay channels for photo-excited electrons in the present experiment. The blue and green arrows correspond to optical transitions (photon energies of 2.09~eV and 3.08~eV, respectively), the black arrows indicate electron relaxation within the band structure and the orange lines represent an Auger process (see text). The red, dotted line represents experimental data obtained from photoemission.\cite{Ast2004}\label{fig:Fig2-ElectronRelaxation}}
\end{figure}

In order to understand the relaxation processes which occur after optical excitation of the electronic system, let us consider the band structure of Bi which is shown in Fig. \ref{fig:Fig2-ElectronRelaxation}. The displayed bands result from calculations by Timrov \emph{et al.}\cite{Timrov2012} which have proven to agree well with 2PPE measurements of the unoccupied band structure\cite{Bronner2013} as well as photoemission experiments of the occupied $p$-bands 3-5 along the $\Gamma T$-line (as indicated in the figure).\cite{Ast2004}

In a first step, an optical transition creates an electron-hole pair. Besides the experiment shown above where the excitation energy was $h\nu=2.09~\text{eV}$, we have conducted an analogous experiment with identical results using a photon energy of 3.08~eV. Both cases are indicated by the blue and green arrows, respectively, in Fig. \ref{fig:Fig2-ElectronRelaxation}. A transition in which electrons are excited into band 6 and to an energy higher than the one of D' can occur in two areas of the Brillouin zone, namely between the $K$- and the $\Gamma$-point as well as between the $\Gamma$ and the $X$-point. However, relaxation of electrons from both regions will lead toward the D' feature at the $\Gamma$-point and therefore we do not need to distinguish between both cases (the corresponding holes decay toward $\Gamma$ in both cases, too).

The transient population of D' occurs within less than 50~fs after optical excitation which is consistent with a relaxation of electrons within band 6. Energetically possible photo-induced transitions of electrons into higher lying bands (i.e. 7 or 8) would be accompanied by subsequent interband transitions into band 6, probably leading to a delayed population of D'. Similarly to this first relaxation, electrons decay along band 6 from D' toward the $T$-point, i.e. into the region associated with feature R. This continuous de-excitation along band 6 is consistent with the short lifetime of the D' feature (less than 125~fs) and the population of R which is observed to occur after the optical excitation and after the population of D' (see inset in Fig. \ref{fig:Fig1-TR2PPE}a). Region R coincides with a true local minimum of the conduction band 6\cite{Timrov2012} and is located at the same point in the Brillouin zone as the hole pocket, with an energetic difference of 330~meV. The relatively long lifetime of feature R, namely 687~fs, is thus consistent with an electronic interband transition from R into the hole pocket.

A population of the hole pocket (region S) by electrons from the local minimum of band 6 (region R) with an inverse rate of 687~fs should also be reflected in the population dynamics of feature S, i.e. the intensity of S should build up on this timescale. However, as evident from our experiments (see inset of Fig. \ref{fig:Fig1-TR2PPE}a), S reaches its highest intensity within less than 200~fs after the excitation. There must hence be another pathway for the decay of excited electrons into the hole pocket. Considering the local band structure at the $T$-point, a direct excitation from band 3 into the hole pocket could be induced by the pump pulse ($h\nu=2.09~\text{eV}$). However, since we observe the same population dynamics also when using a pump photon energy of 3.08~eV, there must be another mechanism. Besides the bulk band structure, electronic relaxation could also occur via surface states and resonances which account for a majority of the density of states in the region of the Fermi level of Bi. As a matter of fact, using time-resolved 2PPE, electrons excited with a pump pulse of 1.6~eV have been observed to decay into an unoccupied branch of the surface band along the $\bar\Gamma\bar M$-line of the surface Brillouin zone.\cite{Faure2013} However, if the predominant decay pathway into the bulk hole pocket was through electron relaxation via the surface band, one would expect a drastic decrease of the intensity as well as a delayed population of feature S when the surface is covered with an adsorbate. However, we observe identical intensities and dynamics for both features R and S in an experiment in which a full monolayer of a large organic molecule (di-meta-cyano-azobenzene\cite{Bronner2013a}) is adsorbed on the Bi(111) surface, even though this quenches the surface electronic structure to a large extent. We can thus exclude also this decay mechanism and instead propose that the hole pocket is populated via an Auger process which occurs in the bulk band structure. During the relaxation of electrons along band 6 and the simultaneous hole decay toward the Fermi level (as discussed below), electron-hole pairs accumulate at the $\Gamma$-point. The energetic difference between the charge carriers is very similar to the energy difference of band 4 and the hole pocket at the $T$-point (see Fig. \ref{fig:Fig2-ElectronRelaxation}). Electron-hole recombination at the $\Gamma$-point could therefore lead to excitation of electrons into the hole pocket. This Auger process could well occur on timescales of a few hundred femtoseconds after the laser pulse, which would explain the relatively fast population of feature S. At the same time, electronic interband transitions from band 6 (region R) into the hole pocket might still occur with an inverse rate of 687~fs. However, the fact that we do not observe an increase of intensity of feature S on this timescale suggests that the Auger process is the predominant pathway.

\subsection{Hole relaxation}

\begin{figure}[htbp]
\includegraphics{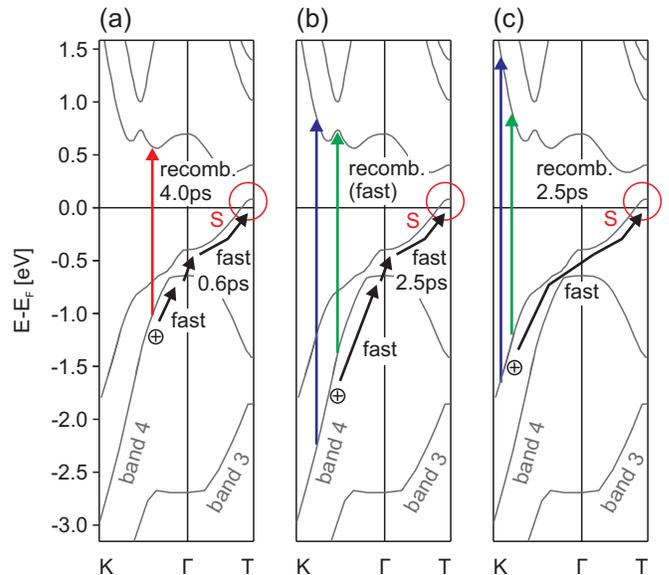}
\caption{(Color online) Decay pathways for the photo-excited hole. (a) Mechanism according to ref. \cite{Timrov2012} upon excitation with $h\nu=1.6~\text{eV}$ (red arrow): The rate-determining step is recombination of the electron-hole plasma. (b) Mechanism in which interband transition dominates the kinetics upon optical excitation with photon energies of 2.09~eV or 3.08~eV, respectively. (c) Mechanism upon excitation with the same photon energies from band 5 with subsequent recombination on the picosecond timescale.\label{fig:Fig3-HoleRelaxation}}
\end{figure}

The initial optical transition does not only create an excited electron but also a hole, which decays toward the Fermi level within a strongly dispersive occupied $p$-band. Let us assume for a moment that the hole is initially created in band 4 as indicated in Fig. \ref{fig:Fig2-ElectronRelaxation}. The photo-hole decays toward the true local maximum\cite{Timrov2012} of band 4 at the $\Gamma$-point where a transition into band 5 is necessary before the relaxation continues along the $\Gamma T$-line. Ultimately, electron and hole recombine at the hole pocket. This mechanism has also been observed in time-resolved THz spectroscopy measurements after excitation of Bi with a photon energy of 1.6~eV, which is shown in Fig. \ref{fig:Fig3-HoleRelaxation}a.\cite{Timrov2012} However, another possible scenario in analogy to the above described decay of the photo-excited electron would be a quasi-one-step decay mechanism of the photo-hole which is governed by the interband transition from band 4 into band 5 (see Fig. \ref{fig:Fig3-HoleRelaxation}b). Furthermore, other optical transitions are energetically possible in the band structure which would lead to the creation of a hole not in band 4 but in band 5 (see Fig. \ref{fig:Fig3-HoleRelaxation}c).

Concerning the decay of the photo-hole, the question is which is the rate-determining step of the electron-hole recombination leading to the observed lifetime of the electrons in the hole pocket of 2.5~ps: either the interband transition from band 4 to band 5 or the recombination of electrons and holes directly at the Fermi level near the hole pocket. Timrov \emph{et al.}\cite{Timrov2012} suggest that the recombination is much slower (with an inverse rate of 4.0~ps) than the interband transition (0.6~ps) due to the low phase space for electron-electron scattering caused by the small Fermi surface of semi-metallic Bi. Despite the difference of the inverse reaction rate (our experiment: 2.5~ps) which might be due to the different excitation energy, this decay mechanism is not in conflict with our time-resolved 2PPE experiments. Since the inverse rate of the fast process in the THz experiment is very similar to the lifetime that we observe for feature R, it is conceivable that this fast process could be interpreted as the interband transition of the electron from band 6 to band 5 as described above. In that case the rate-determining process could well be the interband transition of the photo-hole from band 4 to band 5 (Fig. \ref{fig:Fig3-HoleRelaxation}b), and this transition would have an inverse rate of 2.5~ps. Finally there might be a third possible scenario in which the rate-determining step is indeed the electron-hole recombination at the hole pocket, as suggested by Timrov \emph{et al.}, but which does not involve any interband transition of the hole at all, namely if the hole is initially created in band 5. In this case, the hole could likely decay relatively fast toward the Fermi level. Based on the available experimental results we are unable to rule out any of the three possible mechanisms which we have just discussed and which all match the observations. We would furthermore emphasize that it is well possible that two or all of them occur simultaneously.

One major difference between the three scenarios shown in Fig. \ref{fig:Fig3-HoleRelaxation} is the underlying de-excitation mechanism of the charge carriers at the Fermi level, i.e. the recombination process. Generally, electrons in the hole pocket and holes in the occupied part of band 5 can recombine by coupling to phonons (carrier-phonon scattering) or they can couple to other electrons or holes (carrier-carrier scattering) if there is sufficient phase space. In the decay mechanisms depicted in Fig. \ref{fig:Fig3-HoleRelaxation}a and Fig. \ref{fig:Fig3-HoleRelaxation}c, the electron accumulation in the hole pocket is accompanied by hole buildup in the occupied parts of band 5 which decreases the phase space for carrier-carrier scattering and therefore, relatively slow recombination via carrier-phonon scattering on the picosecond timescale can be expected. The situation shown in Fig. \ref{fig:Fig3-HoleRelaxation}b is different since the holes accumulate in the maximum of band 4 at the $\Gamma$-point. Holes which transition into band 5 and relax toward the $T$-point can recombine quickly with the accumulated electrons in the hole pocket because there is still sufficient phase space for fast carrier-carrier scattering available in this situation.

Note that the above discussion focusses on the charge carrier dynamics which lead to a recombination at the hole pocket at the $T$-point. However, other recombination channels are conceivable, both involving the bulk and the surface electronic structure. For example, relaxation of excited electrons from D' along the $\Gamma L$-line and recombination with the corresponding holes at the $L$-point is very likely. Furthermore, as mentioned above, photo-excited electrons accumulate in the sixfold surface hole pockets along the $\bar\Gamma\bar M$-lines of the surface Brillouin zone while there are holes in the electron pocket of the surface band at the $\bar M$-point. This charge redistribution leads to the excitation of a coherent $A_{1g}$ phonon which oscillates along the [111] direction, i.e. along the $\Gamma T$-line.\cite{{Ohtsubo2012a},{Faure2013}}

As the charge carriers recombine, the coherent phonon is damped with a time constant of 2.6~ps as is observable by following the changes in binding energy of bulk states of Bi.\cite{Papalazarou2012} Within the errors, this value is identical to the lifetime of the electrons in the hole pocket which we observe. It is therefore likely that the transient charge displacement of electrons in the hole pocket and holes in the occupied part of band 5 along the $\Gamma T$-line, leads to the excitation of a coherent $A_{1g}$ phonon in our experiments as well. This would not be surprising since the excitation mechanism of the electronic system is the same but only the photon energy is different.

In contrast to the other 2PPE experiments reported in the literature\cite{{Papalazarou2012},{Faure2013}}, we do not observe an oscillation of the bulk band binding energies upon excitation of the coherent phonon. This should specifically be the case for features D', R and S. There are two major differences to the present experiments: (i) we use a higher photon energy than the 1.6~eV reported in the other studies and (ii) in contrast to the other 2PPE experiments, we observe the electron dynamics along the $\Gamma T$-line, i.e. along the [111] direction. It is conceivable that the changes of the bulk band energies along the $\Gamma T$-line due to the lattice vibrations of the $A_{1g}$ mode are smaller than in other points in the Brillouin zone. Indeed, Faure \emph{et al.}\cite{Faure2013} report on calculations which demonstrate that the effect of atomic displacement on the Bi band energies varies considerably in reciprocal space.

\subsection{Energetic stabilization}

\begin{figure}[htbp]
\includegraphics{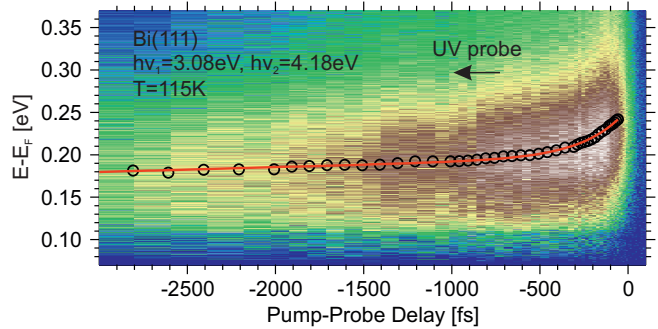}
\caption{(Color online) Energetic stabilization of peak S upon optical excitation. The peak position is indicated by black circles for each delay together with a fit (see text) which describes the stabilization. The color code depicts the measured 2PPE intensity on a linear scale, representing the total population of S.\label{fig:Fig4-Stabilization}}
\end{figure}

Besides the lifetime $\tau_{S}$, we can also observe another dynamic component of state S (see Fig. \ref{fig:Fig4-Stabilization}), namely a continuous decrease of binding energy with time. This energetic stabilization can be described with a two-component time-dependent function for the binding energy which includes a linear and an exponential term,
\begin{equation}
E(\Delta t)-E_{\rm{F}}=a+b\cdot\Delta t+\Delta E\cdot \exp\left({-\frac{\Delta t-t_{0}}{\theta}}\right),
\end{equation}
where $a$, $b$ and $t_{0}$ are fitting constants, $\Delta t$ is the pump-probe delay, $\Delta E$ is the stabilization energy and $\theta$ is the stabilization time constant. The linear term is empirical and probably reflects an additional stabilization process on a longer timescale. The stabilization energy is determined to be $\Delta E=45\pm10~{\rm{meV}}$ which occurs with a time constant of $\theta=209\pm69~{\rm{fs}}$. This energetic relaxation might be due to the formation of an exciton at the $T$-point upon relaxation of the electrons and holes, respectively, to the Fermi level. Due to the low density of states of the semi-metal Bi near the Fermi level, screening of the charge carriers can be expected to be less effective than in metals which results in possible exciton formation. Considering the various decay mechanisms that we have discussed for the hole, this exciton formation within a few hundred femtoseconds would only be consistent with mechanism (c) in Fig. \ref{fig:Fig3-HoleRelaxation}. However, as we have noted, it is possible that more than one hole relaxation mechanisms occur in parallel. Another possible explanation for the energy stabilization is a relaxation of electrons along band 5 within the hole pocket. This relaxation would be independent of the hole decay dynamics.

\section{Conclusion}

Using 2PPE in normal emission we were able to directly probe the population and decay dynamics of the unoccupied electronic states of Bi. The measured lifetimes of the first two IPS are in agreement with previous measurements. More importantly, we can observe the dynamics of (the lowest unoccupied) band 6 at two points in the Brillouin zone as well as the dynamics of electrons in the bulk hole pocket. From the measured dynamics at these three important points in the band structure we can derive a conclusive picture of the relaxation mechanism of photo-excited electrons along the $\Gamma T$-line. One de-excitation channel involves an Auger process which "bypasses" a slow interband transition into the hole pocket where we observe an energetic stabilization on the ultrashort timescale. Furthermore, based on the population dynamics of the electrons in the hole pocket we are able to discuss different mechanism for the relaxation of the photo-hole toward the bulk hole pocket.

2PPE allows to directly follow the charge carrier dynamics of photo-excited Bi. These dynamics have not been investigated before along the $\Gamma T$-line. This region of the Brillouin zone is however particularly interesting since it allows to follow the dynamics at an earlier time after excitation compared to other 2PPE studies which report on the dynamics of relaxed electrons and holes in other parts of the bulk Brillouin zone or in the surface states and resonances. Together with these other investigations, our results contribute to a detailed understanding of the charge carrier relaxation pathways in photo-excited bismuth.

\section{Acknowledgments}
We would like to thank Tobias Kampfrath (Fritz-Haber-Institut Berlin) for fruitful discussions and gratefully acknowledge funding through the Collaborative Research Center (Sfb 658) of the German Research Foundation (DFG).

\end{document}